\documentclass[twocolumn,usenatbib]{mn2e}
\usepackage{amsfonts}
\usepackage{amsmath}
\usepackage{amssymb}
\usepackage{aas_macros}
\usepackage{graphicx}
\usepackage{color}
\usepackage[draft]{hyperref}
\usepackage{float}

\newcommand{\be}{\begin{equation}}
\newcommand{\ee}{\end{equation}}
\def\lsim{\mathrel{\rlap{\lower4pt\hbox{\hskip0.5pt$\sim$}}
    \raise1pt\hbox{$<$}}}         %less than or approx. symbol
\def\gsim{\mathrel{\rlap{\lower4pt\hbox{\hskip0.5pt$\sim$}}
    \raise1pt\hbox{$>$}}}         %greater than or approx. symbol

\def\la{\langle}
\def\ra{\rangle}

\def\mathbi#1{\textbf{\em #1}}

\def\lsim{~\rlap{$<$}{\lower 1.0ex\hbox{$\sim$}}}
\def\bsim{~\rlap{$>$}{\lower 1.0ex\hbox{$\sim$}}}
\def\kms{\ {\rm km\,s^{-1}}}

\def\hkpc{\ {\rm {\it h}^{-1}Kpc}}

\def\kpc{\ {\rm Kpc}}

\def\pc{\ {\rm pc}}

\def\msun{\ {\rm M_\odot}}

\def\msunpc{\ {\rm M_\odot pc^{-3}}}

\def\yr{\ {\rm yr}}
\def\eV{\ {\rm eV}}
\def\GeV{\ {\rm GeV}}

\def\msun{\ M_\odot}

\def\mp{m_\text{P}}
\def\ldb{\lambda_\text{dB}}
\def\vesc{v_\text{esc}}

\def\Mcmin{M_{c,\text{min}}}
\def\Mcmax{M_{c,\text{max}}}
\def\dvir{\Delta_\text{vir}}
\def\meff{m_\text{eff}}

\def\vu{\mathbi{u}}

\def\apjs{{Astrophys.~ J.~ Suppl.~}}
\def\apjl{{Astrophys.~ J.~ Lett.~}}

\definecolor{RedWine}{rgb}{0.743,0,0}
\definecolor{RoyalBlue}{rgb}{0.25,.41,.88}
\definecolor{ForestGreen}{rgb}{.13,.54,.13}
\definecolor{DeepPurple}{rgb}{.72,.18,1}
\begin{document}

\title[Axion core - halo mass and the black hole--halo mass relation: constraints on a few parsec scales]
      {Axion core--halo mass and the black hole--halo mass relation: constraints on a few parsec scales}

\author[Vincent Desjacques and Adi Nusser]
{\parbox[t]{\textwidth}{Vincent Desjacques\thanks{Email: dvince@physics.technion.ac.il} and Adi Nusser}\\\,\\
Physics department and Asher Space Science Institute, Technion, Haifa 3200003, Israel}

\date{}
\label{firstpage}
\pagerange{\pageref{firstpage}--\pageref{lastpage}}

\maketitle

\begin{abstract}
  If  the dark matter is made of ultra-light axions, stable solitonic cores form at the centers of virialized halos. In some range for the mass $m$ of the axion particle,
  these cores  are sufficiently  compact and can mimic  supermassive black holes (SMBH) residing at galactic nuclei. We use the solitonic core--halo mass relation,
  validated in numerical simulations, to constrain a new  range of allowed axion mass from measurements of the SMBH mass in (pseudo)bulge and bulgeless galaxies.
  These limits are based on observations of galactic nuclei on scales smaller than  10 pc.
  Our analysis suggests that $m\lesssim 10^{-18}\eV$ is ruled out by the data.
  We briefly discuss whether an attractive self-interaction among axions could alleviate this constraint.
\end{abstract}


%%%%%%%%%%%%%%%%%%%%%%%%%%%%%%%%%%%%%%%%%%%%%%%%%%%%%%%%%%%%%%
\section{Introduction}
%%%%%%%%%%%%%%%%%%%%%%%%%%%%%%%%%%%%%%%%%%%%%%%%%%%%%%%%%%%%%%

\label{sec:intro}

In the fuzzy dark matter (FDM) scenario
\citep[see, e.g.,][and references therein]{baldeschi/etal:1983,khlopov/etal:1985,sin:1992,hu/etal:2000,svrcek/witten:2006,amendola/barbieri:2006,chavanis:2011,marsh/silk:2014,hlozek/etal:2015,hui/etal:2017},
a halo is made of a solitonic core engulfed by a haze of fluctuating density granules resulting from the interference of (classical) waves.
When the FDM is  ultra-light axions \citep[see, e.g.,][for a recent review]{marsh:2016},  the solitonic core is dubbed ``axion star'' or, simply, an axion core.
Numerical simulations of the Gross-Pitaievskii-Poisson (GPP) system have established that the mass of the axion core $M_c$ increases with the FDM halo mass $M_h$
\citep{schive/etal:2014a,schive/etal:2014b,schwabe/etal:2016}.
Furthermore, simulations have robustly  demonstrated the existence  of a haze of fluctuating granules  extending  much farther than the embedded solitonic core
\citep{schive/etal:2014b}.
This quasi-particle picture has been explored further in \cite{hui/etal:2017,baror/etal:2018} in the context of dynamical friction.
It can also be used to understand the properties of the axion cores.

Measurements from the Lyman-$\alpha$ forest power spectrum set a lower bound on the axion mass of $m\gtrsim 2\times 10^{-21}\eV$ at 95\% C.L.
\citep{irsic/etal:2017,armengaud/etal:2017}.
While our own galaxy could still harbour a solitonic core for a axion mass as low as $m\sim 10^{-22}\eV$ \citep{demartino/etal:2018,broadhurst/etal:2019}, this is quite
unlikely in light of the large scale structure constraints \citep[see, however,][]{zhang/etal:2017}.
The existence of solitonic cores and, thereby, FDM scenarios can be further constrained using different astrophysical observables
\cite[see][for a detailed overview]{hui/etal:2017}, such as galactic rotation curves \citep{bar/etal:2018,robles/etal:2018}, the survival of star clusters in dwarf galaxies
\cite{marsh/niemeyer:2018} or, even, the absence of black-hole superradiance in M87 \citep{davoudiasl/denton:2019}.

Here, we assess the extent to which the presence or absence of supermassive black holes (SMBHs) constrain FDM scenarios. The paper is organized as follows.
After a brief digression on the origin of the axion core -- halo mass relation (\S\ref{sec:axioncore}), we demonstrate that measurements of SMBH and host halo mass in
bulge and, in particular, bulgeless galaxies yield constraints at least as competitive as rotation curves (\S\ref{sec:constraints}). We conclude in \S\ref{sec:conclusions}.

In all illustrations, we assume a concordance $\Lambda$CDM cosmology with Hubble parameter $h=0.7$ and present-day matter density $\Omega_m=0.3$.

%%%%%%%%%%%%%%%%%%%%%%%%%%%%%%%%%%%%%%%%%%%%%%%%%%%%%%%%%%%%%%
\section{Axion core - halo mass relation}
%%%%%%%%%%%%%%%%%%%%%%%%%%%%%%%%%%%%%%%%%%%%%%%%%%%%%%%%%%%%%%

\label{sec:axioncore}

For sake of completeness, we shall discuss briefly the  origin of the axion core--halo mass relation in the context of virial equilibrium, and illustrate
how it can be extended to a non-vanishing (attractive) self-interaction.
More thorough discussions can be found in \cite{chavanis:2011,schive/etal:2014b,marsh/pop:2015,hui/etal:2017}.

We use natural units $c=\hbar=1$ throughout and write Newton's gravitational constant as $G=1/\mp^2$, where
$\mp=1.22\times 10^{19}\GeV$ is the Planck mass.
Furthermore, we parametrize the axion mass $m$ and decay constant $f$ as 
\begin{align}
    m &=10^{-22} m_{22} \eV \\
    f &= 10^{17} f_{17} \GeV \;.
\end{align}     
As a rule of thumb, the present-day axion energy density is given by $\Omega \sim 0.1 f_{17}^2 m_{22}^{1/2}$ \citep{marsh:2016}. Note also that $f_{17}$
quantifies the strength of the axion self-interaction, which we assume attractive .
Since we will consider $f<\mp$ always, we are in the ``strong regime'' of axion self-interactions \citep{chavanis:2017}.

\subsection{Virial equilibrium considerations}

Equilibrum configurations of FDM halos with a density and velocity profile ($\rho$,$\vu$) can be obtained by means of  minimizing
the total energy \citep{chavanis:2011}
\be
E=W + K + Q + U \;,
\ee
where $K$ and $Q$ are the kinetic and the ``quantum pressure'' contributions, $W$ is the gravitational binding
energy of the self-gravitating FDM halo, and $U$ is the ``internal energy'' arising from the self-interaction.
Equilibrium configurations also satisfy the quantum analog of the classical virial theorem \citep{chavanis:2011,hui/etal:2017} implying, in the steady
state limit, 
\be
0 =W + 2K + 2Q +2U \;.
\ee
Since $K\geq 0$, this yields  the inequality $U+Q\leq |W|/2$, which is saturated in the axion core where $K=0$.
By contrast, in the gaseous atmosphere of quasi-particles \citep[see][Appendix \ref{app:quasiparticles}]{hui/etal:2017}, the quantum pressure and
  the self-interaction can be neglected, so that we recover the usual virial theorem $W+2K=0$.

For virialized FDM halos, the velocity dispersion of the quasi-particles surrounding the core is
\be
\label{eq:cond}
\la v^2\ra \approx  \frac{GM_h}{R_h}\; .
\ee
As we shall see now, the core properties are  determined through the requirement that the quasi-particles are barely bound to the core, that is
\begin{equation}
  \label{eq:v2}
\la v^2\ra \approx v_\text{esc}^2 \;,
\end{equation}
where $v_\text{esc}$ is the escape velocity from the axion core.

\subsection{Without self-interaction}

The axion core is characterized by an approximately Gaussian density profile, which reaches a constant central density $\rho_c$ on scales less than
the core radius $R_c$.
In the absence of a self-interaction, $U=0$ and the virial equilibrium condition $W+2Q=0$ inside the solitonic core yields $R_c\propto M_c^{-1}$.
This scaling arises from $W\propto M_c^2/R_c$ and $Q\propto M_c/R_c^2$.
A more detailed analysis gives \citep{chavanis:2011}
\begin{align}
\label{eq:Rcnoself}
R_c &=\frac{3\sqrt{\pi}}{2 M_c}\left(\frac{\mp}{m}\right)^2 \\
&\simeq 227\, m_{22}^{-2}\left(\frac{10^9\!\!\msun}{M_c}\right)\pc 
\nonumber \;.
\end{align}
As a result, the escape velocity $\vesc$ at the surface of the solitonic core is given by
\begin{align}
\label{eq:vesc}
\vesc &= \sqrt{\frac{GM_c}{R_c}} = \sqrt{\frac{2}{3\sqrt{\pi}}}\frac{m}{\mp^2} M_c \\
&\simeq 138\, m_{22}\left(\frac{M_c}{10^9\!\!\msun}\right)\kms
\nonumber \;.
\end{align}
This relation reproduces the empirical scaling $M_c\propto (|E_h|/M_h)^{1/2}$, where $E_h$ is the energy of the halo.
This can also be understood in terms of a wave-like uncertainty principle \citep{schive/etal:2014b}, or diffusive equilibrium \citep{bar/etal:2018}.

The axion core--halo mass relation follows immediately from combining Eqs. (\ref{eq:cond}), (\ref{eq:v2}) and (\ref{eq:vesc}):
\be
\label{eq:mcmh}
M_c = \mathcal{N} \Mcmin^{2/3}\, M_h^{1/3} \;.
\ee
Here, $\Mcmin$ is a minimum core mass,
\begin{align}
\label{eq:mcmin}
  \Mcmin &= \frac{1}{2} 3^{3/4}\pi^{3/8}a^{-3/4}(\Omega_m \dvir)^{1/4} \frac{\mp^2 H_0^{1/2}}{m^{3/2}} \\
  &\simeq 2.51\times 10^7 a^{-3/4} m_{22}^{-3/2} \, \msun \nonumber \;,
\end{align}
and $\mathcal{N}=0.25$ is a empirical normalisation factor which accounts for the fact that the mass assigned to an
axion core in numerical simulations is computed from the central region with $R\lesssim R_c$ only.
The virial overdensity $\dvir(z)$ is defined relative to the average matter density $\bar\rho_m(z)$.
We ignore the mild redshift dependence of $\dvir(z)$ and assume $\dvir(z)=200$ throughout.
Furthermore, $a$ is the scale factor. Since all the data considered here is at redshift $z\ll 1$, we will simply set $a=1$ in all subsequent illustrations.

For a present-day MW-size galaxy with $M_h=4\times 10^{12}\msun$, the axion core mass would be $M_c\sim 5.4\times 10^8 m_{22}^{-1}\msun$.
The minimum core mass $\Mcmin$ originates from the fact that a solitonic core with mass $M_c=\Mcmin$ would have the average density of the Universe
\citep{marsh/pop:2015}. In principle, there is a maximum stable core mass reached when $R_c$ equals the Schwarzschild radius $R_s=2GM_c$.
For realistic CDM cosmologies however, there is not enough time by $z=0$ to form virialized structures which could host axion cores with $M_c\sim M_c(R_c=R_s)$.

Finally, one should bear in mind that, although we will apply Eq.~(\ref{eq:mcmh}) unrestrictedly, it may be a good description of the axion core -- halo
  mass relation solely over a limited range of halo and axion masses \citep[see][for a discussion]{hui/etal:2017}. 

\subsection{With attractive self-interaction}

In the presence of an attractive self-interaction, $U\ne 0$ and the relation between $R_c$ and $M_c$ is more involved. One finds \citep{chavanis:2011}
\be
\label{eq:Rcwself}
R_c = \frac{3\sqrt{\pi}}{4 M_c}\left(\frac{m_\text{P}}{m}\right)^2
\left(1\pm \sqrt{1-\frac{1}{12\pi^2}\left(\frac{m}{m_\text{P} f}\right)^2M_c^2}\right)
\ee
The stable branch corresponds to the plus sign. In this case, the core radius monotonically decreases with increasing
$M_c$ to reach $\frac{3\sqrt{\pi}}{4 M_c}\left(\frac{m_\text{P}}{m}\right)^2$ at the maximum core mass
\begin{align}
  \label{eq:mcmax}
  \Mcmax &= 2\sqrt{3}\pi\left(\frac{m_\text{P} f}{m}\right) \\
  &\simeq 1.19\times 10^{11} \frac{f_{17}}{m_{22}}\msun \nonumber \;,
\end{align}
above which there is no stable solution.

In Appendix \ref{app:quasiparticles}, we show that the quasi-particle approach discussed above also holds in the presence of a self-interaction.
Applying the same hydrostatic considerations yield a core--halo mass relation given by
\be
\label{eq:mcmhwithf}
M_c = \mathcal{N}\, \frac{\Mcmin^{4/3}M_h^{2/3}}{2\Mcmax}\sqrt{\frac{4\Mcmax^2}{\Mcmin^{4/3}M_h^{2/3}}-1}\;.
\ee
For the normalisation, we shall adopt again $\mathcal{N}=0.25$.
The axion core mass reaches its maximum $M_c=\Mcmax$ for a halo mass
\begin{align}
  \tilde{M}_h &= 2^{3/2}\frac{\Mcmax^3}{\Mcmin^2} \\
  &= 7.57\times 10^{18} a^{3/2}f_{17}^3\msun \nonumber 
\end{align}
independently of the axion mass $m_{22}$. 
For $M_h\geq\tilde{M}_h$,  hydrostatic equilibrium cannot be satisfied.
Note that Eq.(\ref{eq:mcmhwithf}) recovers Eq.(\ref{eq:mcmh}) in the limit $f_{17}\to \infty$, that is, in the absence of self-interactions.

An attractive self-interaction lowers the minimum core mass obtained upon setting $M_c=M_h$. 
However, for values of $f_{17} \gtrsim 0.01$ compatible with all axions being the dark matter, this is at most a factor
of 2 smaller than $\Mcmin$: the axion self-interaction scale as $\propto\rho^2$ and, thus, is very weak at low densities.

\subsection{Mergers and the persistence of axion cores}

The equilibrium considerations above do not take into account the evolution of $M_c$ and $M_h$ through mergers and smooth accretion, which is an essential
aspect of hierarchical structure formation.
A related issue is the persistence of the axion core -- halo mass relation Eq.~(\ref{eq:mcmh}) through the assembly history of the host FDM halos
\citep[see, e.g.][]{schwabe/etal:2016,du/etal:2017b}

Although numerical simulations indicate that cores are ubiquitous inside FDM halos \citep{schive/etal:2014a,veltmaat/etal:2018},
the fate of solitonic cores during the merger of two FDM halos is unclear.
Therefore, a lack of evidence for a central core does not necessarily translate into a constraint on the axion mass, unless the
characteristic timescale for the formation of a new core following a merger event is shorter than the age of the galaxy. 

The cores of the progenitor FDM halos may i) remain intact, or ii) momentarily disappear during the merging process.
To determine whether a core forms in the central region of the descendant FDM halo, one should thus consider either i) the dynamical friction
timescale on which they sink to the center of the merged halo, or ii) the relaxation timescale of FDM quasi-particle, which defines the
region within which virial equilibrium can be established.
Furthermore, all this could depend on the axion mass since the solitonic cores become more compact as $m_{22}$ increases
and, therefore, are less likely to be disrupted.
For simplicity however, we will assume that scenario ii) is the relevant picture for the range of axion masses considered in Fig.\ref{fig:mvcirc}.
This scenario likely applies to  major mergers during which the gravitational potential fluctuates significantly on a short timescale and,
thereby, destroys the coherence of the axion core. 

Under this assumption, the relevant timescale is the two-body relaxation timescale between the FDM quasi-particles. As discussed in \cite{hui/etal:2017}
\cite[and][in further details]{baror/etal:2018}, this reads
\begin{equation}
  \label{eq:trelax}
t_\text{relax} = \frac{10^{10}\yr}{f_\text{relax}}m_{22}^3 \left(\frac{v}{100\kms}\right)^2\left(\frac{r}{5\kpc}\right)^4 \;.
\end{equation}
An FDM halo will develop a compact solitonic core from the mass bound to the descendant halo within a radius $R_c$ if the condition
$t_\text{relax}(R_c)\lesssim t_\text{mg}$ is satisfied. Here, $t_\text{mg}$ is the time elasped since the merger. Setting $v=V_\text{circ}$ in the above
expression, and using the core -- halo mass relation Eq.(\ref{eq:mcmh}), the newly merged halo will develop an axion core of mass $M_c\propto M_h^{1/3}$
provided that
\begin{equation}
M_c \gtrsim 3.5 \times 10^4\msun \frac{a^{1/2}m_{22}^{-3/2}}{f_\text{relax}^{1/2}} \left(\frac{10^{10}\yr}{t_\text{mg}}\right)^{1/2} \;.
\end{equation}
Although $t_\text{relax}$ increases with the axion mass, the minimum core mass scales like $M_c\propto m_{22}^{-3/2}$ because of the core radius
$R_c\propto m_{22}^{-2}$ shrinks rapidly as the axion mass is increased. 
Assuming $f_\text{relax}\sim 1$ and $t_\text{mg}=H_0^{-1}$ for illustration, where $H_0$ is the Hubble constant today, this condition is satisfied for
the whole range of circular velocities and masses shown in Fig.\ref{fig:mvcirc}.

Requiring that the whole descendant FDM halo be in virial equilibrium (which amounts to setting $v=V_\text{vir}$ and $r=R_\text{vir}$ in
Eq.(\ref{eq:trelax})) would ensure that the axion core mass of the merged halo precisely falls on the relation Eq.(\ref{eq:mcmh}). However, we found
that such a condition cannot be satisfied unless the core mass is close to $\Mcmin$ (so that the FDM atmosphere is tenuous). Therefore, one should
expect some scatter in $M_c$ at fixed halo mass. 

Note that the sum of the progenitor core mass is always larger than the core mass expected if the final descendant halo reaches hydrostatic equilibrium.
To see this, let $M_{h1}$, $M_{h2}$ be the mass of the progenitor halos, with corresponding core mass $M_{c1}$ and $M_{c2}$; and $M_h=M_{h1}+M_{h2}$ be the
mass of the merged halo. Let us also define $M_c=M_{c1}+M_{c2}$. Assuming that the core - atmosphere of the progenitors is in hydrostatic equilibrium, so
that Eq.(\ref{eq:mcmh}) initially holds, we have 
\be
M_c = \Mcmin^{2/3} M_h^{1/3} \left[1+3\frac{M_{h1}^{2/3}M_{h2}^{1/3}+M_{h1}^{1/3} M_{h2}^{2/3}}{M_h}\right]^{1/3} \;,
\ee
which shows that $M_{c1}+M_{c2}> \Mcmin^{2/3}M_h^{1/3}$. The difference is maximum for a major merger with $M_{h1}\approx M_{h2}$, in which case
$M_c\approx 1.6\,\Mcmin M_h^{1/3}$.

%%%%%%%%%%%%%%%%%%%%%%%%%%%%%%%%%%%%%%%%%%%%%%%%%%%%%%%%%%%%%%
\section{Constraints on axion mass from $M_\bullet$ - $V_\text{circ}$ measurements}
%%%%%%%%%%%%%%%%%%%%%%%%%%%%%%%%%%%%%%%%%%%%%%%%%%%%%%%%%%%%%%

\label{sec:constraints}

We discuss now the constraints on the axion mass $m$ that arise from measurements of the mass, $M_\bullet$, of SMBH residing at galactic nuclei, and from the
galactic  (asymptotic) circular velocity $V_\text{circ}$ at larger radii. The asymptotic 
circular velocity is used as a proxy for the host halo mass $M_h$. The full  rotation curve is irrelevant for the constraints derived here.
For sufficiently small $R_c$, the axion core could masquerade as a galactic SMBH.  \cite{hui/etal:2017} briefly discussed this possibility for large galaxies.
Here, we will show that small galaxies actually give the strongest limits on $m$.

\subsection{Strategy}

Observational constraints on galactic SMBH masses are mainly obtained from studying the stellar kinematics within small distances ($R_e<10\!\pc$)  of a few
times the radius of influence of the  SMBH. When an estimate of the host halo mass $M_h$ is available, the axion core radius $R_c(M_c,m)$ and mass $M_c(M_h,m)$
can be obtained from the  relations (\ref{eq:Rcnoself}) and (\ref{eq:mcmh}), respectively.
More precisely, taking into account the dependence $\Mcmin\propto m^{-1}$, cf. Eq.(\ref{eq:mcmin}), we find
\begin{equation}
    M_c\propto \frac{M_h^{1/3}}{m}\;,\qquad
    R_c\propto \frac{1}{mM_h} \;.
\end{equation}
On the one hand, too low values for $m$ imply large core masses $M_c$, yet constraints cannot be obtained because the core is too diffuse.
On the other hand, too high $m$ cannot be ruled out either since they yield  $M_c\ll M_\bullet$.
Therefore, this technique can constrain a limited, albeit interesting range of $m$ where the core is sufficiently compact and massive.

To compare the data to theoretical expectations based on the axion core -- halo mass relation, we need to associate the observed  circular velocity $V_\text{circ}$
to the  halo mass $M_h$. We adopt the following  relation 
\begin{equation}
  V_\text{circ} \approx 144 \kms \left(\frac{M_h}{10^{12}\msun}\right)^{1/3} \;,
\end{equation}
which assumes an overdensity threshold $\Delta_\text{vir}=200$ (in unit of the critical density $\rho_\text{cr}$). This allows us to convert $M_c(M_h,m_{22})$ into a
relation $M_c$ - $V_\text{circ}$ once an axion mass is assumed.

\subsection{Data}

The  analysis requires a sample of measured black hole masses and circular velocities of  the respective host halos. 
\cite{kormendy/ho:2013} provides an excellent review of the relevant techniques for measuring SMBH masses, as well as a discussion of the correlations between the
inferred masses and properties of their host galaxies. The tightest correlation is between  $M_\bullet$ and the velocity dispersion $\sigma$ of the central stellar
component. 
Fortunately,  \cite{kormendy/ho:2013} also list the circular velocities $V_\text{circ}$  of many of the host galaxies given in their paper.
For spiral galaxies, $V_\text{circ}$ is derived from the rotation curves while, for ellipticals, it is simply $\sqrt{2} \sigma$. At a given galaxy mass, the least
massive SMBH are found in spirals with pseudobulges or no bulge at all.  Thus, we expect that the strongest constraints will be obtained using these galaxies,
rather than ellipticals or galaxies with classical bulges. 
 
In Fig.\ref{fig:mvcirc}, we display measurements from classical and pseudo-bulges as (filled) red and (empty) blue circles, respectively, along with the empirical
relation
\begin{equation}
  M_\bullet \approx 0.32\times 10^8 \msun \left(\frac{V_\text{circ}}{200\kms}\right)^{5.1}
\end{equation}
as the thick black line. The powerlaw scaling reflects the relationship advocated by \cite{McConnell:2011mu} \citep[see also][]{Ferrarese:2000se}.

Furthermore, we display measurements from bulgeless galaxies as the green squares.
Except for NGC 4395, for which a reverberation-mapping measurement gives $M_\bullet=(3.6\pm 1.1)\times 10^5\msun$ \citep{peterson/etal:2005},
all these measurements provides an upper limit on the mass of the SMBH. 
For NGC 300, 3423, 7424 and 7793, the limits on $M_\bullet$ are from \cite{neumayer/walcher:2012}. $V_\text{circ}$ for NGC 7424 is from \cite{sorgho/etal:2019} whereas,
for NGC 7793,  $V_\text{cric}$ is from \cite{deblok/etal:2008}. 
Finally, M33 (a nearby spiral galaxy embedded in a dark matter halo) has an asymptotic circular velocity of $V_\text{circ}\approx 125\kms$ \citep{mayall/aller:1942} and
the tightest upper limit on the SMBH mass: $M_\bullet < 1.5\times 10^3\msun$ \citep{gebhardt/etal:2001}. We have thus labelled the corresponding data point on Fig.~\ref{fig:mvcirc}. The data of the bulgeless galaxies is all summarized in Table \ref{table:table1}.

%The   $M_c$ - $V_\text{circ}$ relations, shown as the dashed lines in Fig.\ref{fig:mvcirc}, assume there is no self-interaction
%($f_{17}=0$). The axion mass increases in steps of an order of magnitude, from $m_{22}=1$ until $m_{22}=10^5$ (from top to bottom).
%By contrast, the two dotted curves are for a fixed axion mass $m_{22}=10^2$, but a decay constant $f_{17}=0.005$ and $0.01$ (from
%left to right). Numerical simulations indicate that, alhough the correlation between $M_c$ and $M_h$ is tight, one should expect a scatter
%of order 0.3 dex, presumably arising from the imperfect relaxation after merger events etc. outlined in \S\ref{sec:axioncore}.

\subsection{Constraints}

\begin{figure*}
  \includegraphics[width=15cm]{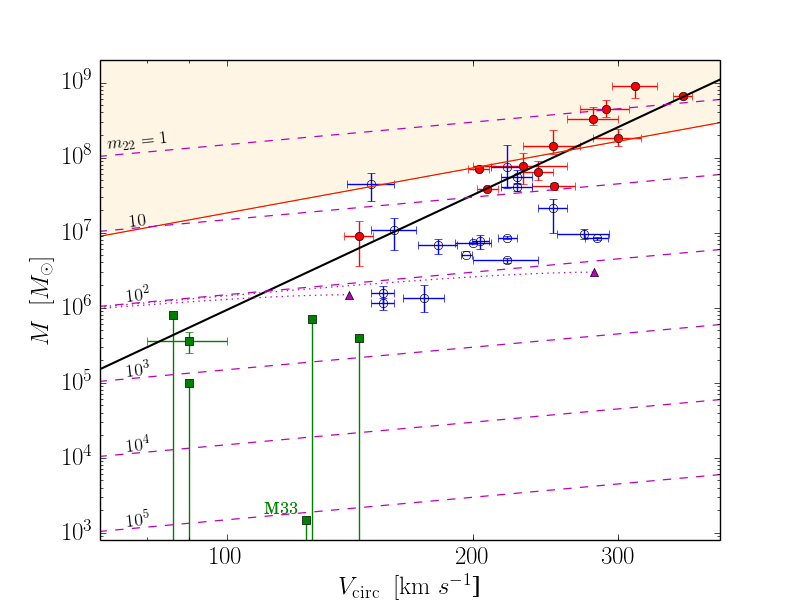}
  \caption{Measurements of central SMBH mass $M_\bullet$ vs circular velocity $V_\text{circ}$ for different types of galaxies.
    Filled red and empty blue circles designate the dynamical measurement of $M_\bullet$ in classical and pseudobulges, respectively, while the
    green squares represent (mostly) upper limits for bulgeless galaxies \citep[from the compilation of][]{kormendy/ho:2013}.
    The galaxy with the tightest black hole mass upper limit is M33 as indicated on the figure.
    The thick black line is the empirical $M_\bullet$ - $V_\text{circ}$ relation, whereas the dashed and dotted curves indicate the axion
    core mass $M_c$ vs. $V_\text{circ}$ expected for ultra-light axions w/o self-interactions (see text for details).
    The shaded orange area shows the region in which the core radius $R_c$ is larger than 10 pc.}
  \label{fig:mvcirc}
\end{figure*}

It is instructive to first compare the data to the axion core -- halo mass relations discussed in \S\ref{sec:axioncore}.
For this purpose, we overlay in Fig.\ref{fig:mvcirc} the $M_c$ - $V_\text{circ}$ relations inferred from Eq.(\ref{eq:mcmh}) (dashed lines,
no self-interaction) and Eq.(\ref{eq:mcmhwithf}) (dotted lines, with self-interactions). The axion mass increases in steps of an order of
magnitude, from $m_{22}=1$ until $m_{22}=10^5$ (from top to bottom). The effect of a self-interaction is shown only for a mass $m_{22}=10^2$.
The two dotted curves assume a decay constant $f_{17}=0.005$ and $0.01$ (from left to right). The triangle marks the value of $V_\text{circ}$
at which $M_c=\Mcmax$, where $\Mcmax$ is given by Eq.~(\ref{eq:mcmax}). We have not shown the scatter expected around the $M_c$ - $V_\text{circ}$
relation owing to the imperfect relaxation after merger events etc. outlined in \S\ref{sec:axioncore}. Numerical simulations indicate that this
scatter is of order 0.3 dex for halo masses $10^9\lesssim M_h\lesssim 10^{11}\msun$ \citep{schive/etal:2014b}.

%The   $M_c$ - $V_\text{circ}$ relations, shown as the dashed lines in Fig.\ref{fig:mvcirc}, assume there is no self-interaction
%($f_{17}=0$). The axion mass increases in steps of an order of magnitude, from $m_{22}=1$ until $m_{22}=10^5$ (from top to bottom).
%By contrast, the two dotted curves are for a fixed axion mass $m_{22}=10^2$, but a decay constant $f_{17}=0.005$ and $0.01$ (from
%left to right). Numerical simulations indicate that, alhough the correlation between $M_c$ and $M_h$ is tight, one should expect a scatter
%of order 0.3 dex, presumably arising from the imperfect relaxation after merger events etc. outlined in \S\ref{sec:axioncore}. 

Axion cores could mimic a point source like a SMBH provided their radius is smaller than the radius $R_e$ of the central nuclear star cluster,
the velocity dispersion of which constrains the SMBH mass. Typical values of $R_e$ are in the range $R_e \sim 1 - 10 \pc$.
For illustration, the thick orange line shows the locus $M_c(m_{22})$ for which the core radius is $R_c=10\ {\rm pc}$, so that the shaded
upper half of Fig.\ref{fig:mvcirc} corresponds to axion cores with a radius $R_c>10\ {\rm pc}$. Such axion cores cannot be approximated as a
central point source similar to a SMBH.
This excludes the possibility that the classical bulges with $V_\text{circ}\gtrsim 250\kms$ actually harbor axion cores.

%We now turn on the constraints that can be set on the axion mass $m$.

Measurements of the Lyman-$\alpha$ forest rule out the range $m\lesssim 2\times 10^{-21}\eV$ (i.e. $m_{22}\lesssim 20$) at 95\% C.L.
\citep{irsic/etal:2017,armengaud/etal:2017}.
Such low values of $m$ yield an axion core -- halo mass relation (at least partly) in the orange shaded region of Fig.\ref{fig:mvcirc} and, therefore,
typically correspond to large core radii $R_c$ which cannot mimic a central point source.
Furthermore, if the hypothetical axion has a mass $m_{22}\gtrsim 20$, then essentially all the classical bulges must correspond to SMBHs.
Fig.\ref{fig:mvcirc} also suggests that, if this axion  would be self-interacting with a decay constant $0.005\lesssim f_{17}\lesssim 0.01$, the
low mass compact objects harbored mainly by pseudobulges could actually be axion cores, while the more massive ones would have exceeded the threshold
Eq.(\ref{eq:mcmax}) and collapsed to form black holes. In such a scenario, two populations of central compact objects - axion cores and SMBHs - could
coexist over a certain range of $V_\text{circ}$.
However, our discussion thus far does not take into account the non-detection of central compact objects in nearly all the low-$V_\text{circ}$
bulgeless galaxies. This yields the strongest constraints on $m$ as we shall see now. 

To exclude a range of axion mass from bulgeless galaxies, we require that the core mass within the radius of the nuclear cluster be less than the
maximum black hole mass inferred from the nuclear star cluster. 
More precisely, let $R_e$ be the radius of the central stellar cluster. 
There are two possibilities depending on
whether $R_c$ is larger or smaller than $R_e$.
If $R_c>R_e$, we demand 
\begin{equation}
\label{eq:request1}
M_c \left(\frac{R_e}{R_c}\right)^3 < M_{\bullet,\text{max}} \;,
\end{equation}
where $M_{\bullet,\text{max}}$ is the upper limit on the SMBH mass as given in Table~\ref{table:table1}. 
Assuming the core is in hydrostatic equilibrium so that relation Eq.(\ref{eq:mcmh}) holds,
we can express both $M_c$ and $R_c$ as a function of $M_h$ or, equivalently, $V_\text{circ}$.
As a result, Eq.~(\ref{eq:request1}) translates into an upper limit on the allowed axion mass of
\begin{align}
m_{22} &\lesssim 9.7\,\frac{a}{\sqrt{\mathcal{N}}} \left(\frac{M_{\bullet,\text{max}}}{10^3\msun}\right)^{1/2}\left(\frac{1\pc}{R_e}\right)^{3/2} \\
&\qquad\times
\left(\frac{100\kms}{V_\text{circ}}\right)^2 
\nonumber \;,
\end{align}
in which we set $a=1$ and $\mathcal{N}=0.25$ as advocated above.
The multiplicative term $\sqrt{\mathcal{N}}$ arises from the fact that, in Eq.(\ref{eq:request1}), $M_c$ comes with one normalisation factor $\mathcal{N}$
(since $M_c$ represents the mass enclosed within $R_c$ solely), while $R_c$ does not.

If $R_c<R_e$, then the core mass must satisfy
\begin{equation}
    \frac{3\sqrt{\pi}}{2}\left(\frac{m_\text{P}}{m}\right)^2\frac{1}{R_e}<M_c<M_{\bullet,\text{max}}\;.
\end{equation}
This translates into a lower limit on the allowed axion mass of
\begin{equation}
m_{22} \gtrsim 1.5\times 10^4 \left(\frac{10^3\msun}{M_{\bullet,\text{max}}}\right)^{1/2}\left(\frac{1\pc}{R_e}\right)^{1/2}
\end{equation}
independently of the host halo mass. 

Values of $R_e$ are obtained from \cite{gebhardt/etal:2001} for M33, and from \cite{neumayer/walcher:2012} for the remaining galaxies.
They are all summarized in Table~\ref{table:table1}, along with the constraints on $m$.
Taking into account the finite extent of the nuclear cluster, the actual limits on the axion mass are different from those directly inferred from
Fig.~\ref{fig:mvcirc} (e.g. we would read off $m_{22}\gtrsim 10^5$ from M33).
Notwithstanding, the range of low axion masses allowed by this data, $m\lesssim 10^{-21}\eV$, is incompatible with the Lyman-$\alpha$ forest constraints.
Therefore, if dark matter is an ultra-light axion, then its mass must exceed the lower limits given in the last column of Table~\ref{table:table1} in order
to satisfy the constraints from bulgeless galaxies.
The absence of a compact object at the center of M33 gives the strongest constraint: $m>1.2\times 10^{-18}\eV$ (or, equivalently, $m_{22}>1.2\times 10^4$).

%%%%%%%%%%%%%%%%%%%%%%%%%%%%%%%%%%
\begin{table*}
  \caption{Constraint on axion mass from bulgeless galaxies. $M_\bullet$ is the mass of the central SMBH (in $\msun$),
    $R_e$ is the radius of the nuclear star cluster (in $\pc$), $V_\text{circ}$ is the asymptotic circular velocity (in $\kms$), and the axion
    mass $m_{22}$ is in unit of $10^{-22}\eV$. The constraints on the axion mass assume that the core radius $R_c$ is either larger (left
    column) or smaller (right column) than $R_e$. See text for details.
    }
\begin{center}
  \begin{tabular}{lcccll} \hline\hline
    &  $M_\bullet$ & $R_e$ & $V_\text{circ}$ & \multicolumn{2}{c}{constraint on $m_{22}$} \\ \hline
  M33 & $< 1.5\times 10^3$ & 1.0 & 125 & $< 15$ & $> 1.2\times 10^4$ \\
  NGC 300 & $< 10^5$ & 2.9 & 90 & $< 48$ & $> 880$ \\
  NGC 3423 & $< 7\times 10^5$ & 4.18 & 127 & $< 36$ & $> 280$ \\
  NGC 4395 & $(3.6\pm 1.1)\times 10^5$ & -- & 90 & -- & -- \\
  NGC 7424 & $< 4\times 10^5$ & 7.4 & 145 & $< 9.0$ & $> 270$ \\
  NGC 7793 & $< 8\times 10^5$ & 7.7 & 86 & $< 34$ & $> 190$ \\ \hline\hline
\end{tabular}
\end{center}
\label{table:table1}
\end{table*}

%%%%%%%%%%%%%%%%%%%%%%%%%%%%%%%%%%%%%%%%%%%%%%%%%%%%%%%%%%%%%%
\section{Conclusions}
%%%%%%%%%%%%%%%%%%%%%%%%%%%%%%%%%%%%%%%%%%%%%%%%%%%%%%%%%%%%%%

\label{sec:conclusions}

We have assessed the extent to which measurements of the SMBH mass $M_\bullet$, and the halo mass $M_h$ for bulge and bulgeless galaxies can prove the mass $m$ of
an hypothetical ultra-light axion dark matter.
This data can constrain an interesting range of axion mass $10^{-20} - 10^{-18}\eV$ for which the axion cores are neither too diffuse nor too massive.

While we have used the compilation of \cite{kormendy/ho:2013} for (pseudo)bulge galaxies, small bulgeless galaxies actually give the strongest constraints on $m$.
In particular, the non-detection of a central compact object in M33 -- an isolated spiral galaxy from the local group without any indication of recent mergers or interactions with other galaxies \citep{verley/etal:2009} -- gives $m\gtrsim 1.2\times 10^{-18}\eV$.
The range of mass $10^{-19} - 10^{-18}\eV$ is not easily accessible to measurements from rotation curves, which typically probe scales $r\gg 1\pc$
\citep[e.g.][]{slepian/goodman:2012,bar/etal:2018}. Our constraints also improve on those inferred by \cite{marsh/niemeyer:2018} from the presence of old star
clusters in Eridanus II.
We stress that our limits rely on the caveat that the axion core -- halo mass relation Eq.(\ref{eq:mcmh}) is valid in a range of axion masses for which it has, in fact, not been tested numerically. Therefore, our constraints would weaken, would the hypothetical axion core mass fall below the relation Eq.~(\ref{eq:mcmh}).

Instability of the axion core owing to an attractive self-interaction \cite[e.g.,][]{chavanis:2011,visinelli/etal:2018} with $f_{17}\lesssim 0.01$
\citep[possibly amplified by external perturbers, cf.][]{eby/etal:2018}, could help relaxing the constraints on $m$ if, during the collapse to a black hole, a
significant fraction of the axion core mass can be expelled. Fig.\ref{fig:mvcirc} shows that, for a galaxy with a mass comparable to M33, an axion decay constant
  $f\lesssim 5\times 10^{14}\GeV$ is required for the harbored axion core to be unstable. It is unclear whether such low values of $f$ could still produce the right
  relic abundance, although temperature-dependent effects during the symmetry breaking can help achieving $\Omega\sim 0.1$ \citep{Diez-Tejedor:2017ivd}. In any case,
  it is pretty clear that such a self-interacting model would have to be somewhat fine-tuned in order to satisfy both M33 and cosmological constraints.

What is the fate of unstable axion cores ? Self-similar solutions to the "wave collapse" indicate that interactions
near the center create an outgoing stream of particles which can carry away a large fraction of the axion core before the formation of a black hole remnant
\citep[see, e.g.,][]{levkov/etal:2017}. It would be interesting to investigate whether this effect can produce a range of remnant SMBH masses broad enough to explain
the non-detection of a SMBH in M33, together with the detection of a $\sim 10^5\msun$ SMBH in NGC 4395.

\section{Acknowledgments}

V.D. acknowledges support by the Israel Science Foundation (grant no. 1395/16).
A.N. acknowledges support by the I-CORE Program of the Planning and Budgeting Committee, the Israel Science Foundation (grants No. 1829/12 and No. 936/18) and the Asher
Space Research Institute.

\appendix

%%%%%%%%%%%%%%%%%%%%%%%%%%%%%%%%%%%%%%%%%%%%%%%%%%%%%%%%%%%%%%
\section{On the quasi-particle description of FDM halos}
%%%%%%%%%%%%%%%%%%%%%%%%%%%%%%%%%%%%%%%%%%%%%%%%%%%%%%%%%%%%%%

\label{app:quasiparticles}

Motivated by numerical simulations, \cite{hui/etal:2017} suggested that the atmosphere of FDM halos can be approximated as a gas of quasi-particles of characteristic size
$\ldb$, where $\ldb=(mv)^{-1}$ is the de Broglie wavelength of the axion particle. For a typical velocity $v\sim 10^{-4}$, $\ldb\sim m_{22}^{-1}\hkpc$ is on galactic scales.

Numerical simulations show that the binding energy of these quasi-particles is negligible compared to their kinetic
energy \citep{veltmaat/etal:2018}. In other words, their self-gravity can be neglected so that their dispersion relation is that of a free particle, $\omega(k) = k^2/2m$.
As a result, a quasi-particle of initial width $\ldb$ gradually spreads over with a rms
width given by $\sqrt{(\ldb^4 + (t/m)^2)}/\ldb$. Quasi-particles thus disperse after a time $\tau\sim \ldb^2/m$, that is,
\be
\tau\sim 2.1\times 10^7\,m_{22}^{-1} \left(\frac{v}{10^{-4}}\right)^2\,\yr \;,
\ee
in agreement with the findings of \cite{veltmaat/etal:2018}.
The number $N$ of quasi-particles populating a FDM halo is, therefore, not conserved. Nevertheless, we expect the average number $\la N\ra$ to be conserved for FDM halos
in virial equilibrium.

Turning on the self-interaction should not affect this picture noticeably.
To see this, we assume that the quasi-particles are described by Gaussian wave packets of size $\ldb$ and mass
$\meff=(2\pi)^{3/2}\rho \ldb^3$ as in \cite{hui/etal:2017}.
Here, $\rho$ is the density in the FDM atmosphere surrounding the axion core.
The various energy contributions straightforwardly follow from the Gaussian ansatz used by \cite{chavanis:2011}.
We find:
\begin{align}
  K&=\frac{1}{2m^2} \frac{\meff}{\ldb^2} \;,\qquad Q = \frac{\sigma_3}{m^2} \frac{\meff}{\ldb^2} \\
  U &= \frac{\zeta_3}{m^2 f^2}\frac{\meff^2}{\ldb^3} \;,\qquad W=\frac{\nu_3}{\mp^2} \frac{\meff^2}{\ldb} \nonumber \;,
\end{align}
where
\be
\sigma_3=\frac{3}{8}\;,\quad \zeta_3=-\frac{1}{128\pi^{3/2}}\;,\quad
\nu_3=-\frac{1}{2\sqrt{\pi}}
\ee
This gives
\be
\bigg\lvert\frac{K}{W}\bigg\lvert \simeq
5.7\times 10^{-2} m_{22}^2 \left(\frac{\rho}{\msunpc}\right)^{-1}\left(\frac{v}{10^{-4}}\right)^4 \;,
\ee
and
\be
\bigg\lvert\frac{K}{U}\bigg\lvert \simeq
2.4 m_{22}^2 f_{17}^2 \left(\frac{\rho}{\msunpc}\right)^{-1}\left(\frac{v}{10^{-4}}\right)^2 \;.
\ee
Note that a relic axion dark matter density $\Omega$ today implies $m_{22}^{1/2}f_{17}^2\sim \Omega/0.1$. Therefore, at fixed $\Omega$, $|K/U|$ only weakly depends on $f_{17}$.

The wave packet behaves as a transient quasi-particle if its kinetic energy is much larger than both its potential and internal energy, $W$ and $U$. For an axion mass
$m_{22}\gtrsim 0.1$, the conditions $|K|\gg |W|$ and $|K|\gg |U|$ are satisfied provided that $\rho\lesssim 1\msunpc$. 
For comparison, the dark matter density in the neighborhood of the solar system is $\rho_\odot\simeq 0.5-1 \times 10^{-2}\msunpc$.

To conclude, note that there is an interesting similarity between the properties of halos in axion dark matter, and in repulsive BEC (Bose-Einstein condensate) dark matter
cosmologies. In the latter case, the mass of the dark matter particle is orders of magnitude larger than $10^{-22}\eV$, so that the delocalization arising from the de Broglie
wavelength is irrelevant. What provides the "pressure" support is a repulsive interaction, rather than the "quantum pressure". Interestingly, dense cores also form at the
centers of virialized halos, which can affect rotation curves \citep{slepian/goodman:2012}.
This suggests it should be possible to adapt the approach of \cite{chavanis:2018} to describe the core and atmosphere of FDM halos.

\bibliographystyle{mn2e}
\bibliography{references}

\label{lastpage}

\end{document}